\documentclass[5p,twocolumn,sort&compress,times,preprint,normalem]{elsarticle}

\usepackage{graphics}
\usepackage{graphicx}

\usepackage{amsfonts, amsthm, amssymb}
\usepackage{color}
\usepackage{ascmac}
\usepackage{url}
\usepackage{ulem}
\usepackage{bm}
\usepackage{bbold}
\usepackage{amsmath}
\usepackage{algorithm}
\usepackage{algpseudocode}
\usepackage{braket}
\usepackage{multirow}
\usepackage{subfiles}
\usepackage{cprotect}
\usepackage{epsfig}
\usepackage{xspace}
\usepackage{tipa}

\newcounter{bla}

\journal{Computer Physics Communications}

\begin{document}

\begin{frontmatter}



\title{Data-driven sensitivity analysis 
in surface structure determination 
using  total-reflection high-energy positron diffraction (TRHEPD)}


\author[a,b,c]{Takeo Hoshi\corref{author}}
\author[a]{Daishiro Sakata}
\author[d]{Shotaro Oie}
\author[c]{Izumi Mochizuki}
\author[d]{Satoru Tanaka}
\author[c]{Toshio Hyodo}
\author[e]{Koji Hukushima}
\cortext[author] {Corresponding author.\\\textit{E-mail address: hoshi@tottori-u.ac.jp} }

\address[a]{Department of Applied Mathematics and Physics, Tottori University,  Tottori-shi, Tottori 680-8552, Japan}
\address[b]{Advanced Mechanical and Electronic System Research Center, Faculty of Engineering, Tottori University, 4-101 Koyama-Minami, Tottori 680-8550, Japan}
\address[c]{Slow Positron Facility, Institute of Materials Structure Science, High Energy Accelerator Research Organization (KEK), Oho 1-1, Tsukuba, Ibaraki, 305-0801, Japan}
\address[d]{Department of Applied Quantum Physics and Nuclear Engineering, Kyushu University, 744 Motooka, Nishi-ku, Fukuoka 819-0395, Japan}
\address[e]{Graduate School of Arts and Sciences, The University of Tokyo 153-8902, Japan}

\begin{abstract}
The present article proposes a data analysis method for experimentally-derived measurements, 
which consists of an auto-optimization procedure and a sensitivity analysis. 
The method was applied to the results of a total-reflection high-energy positron diffraction (TRHEPD) experiment,
a novel technique of determining surface structures or the position of the atoms near the material surface.  
This method solves numerically the partial differential equation in the fully-dynamical quantum diffraction theory with many trial surface structures. In the sensitivity analysis, we focused on the experimental uncertainties and the variation over individual fitting parameters, which was analyzed by solving the eigenvalue problem of the variance-covariance matrix. A modern massively parallel supercomputer was used to complete the analysis within a moderate computational time. The sensitivity analysis provides a basis for the choice of variables in the data analysis for practical reliability. The effectiveness of the present analysis method was demonstrated in the structure determination of 
a Si$_4$O$_5$N$_3$ / 6H-SiC(0001)-($\sqrt{3} \times \sqrt{3}$) R30$^\circ$ surface. 
Furthermore, this analysis method is applicable to many experiments other than TRHEPD.
\end{abstract}


\end{frontmatter}
\section{Introduction}

Data analysis procedures with practical reliability and moderate computational times 
are necessary features for measurement techniques in materials science and other physical fields.
In general, the data analysis procedures determine 
target variables $X \equiv (X_1, X_2, ..., X_n)$  from obtained experiment data, $F_{\rm exp}$, 
for a chosen characteristic $F(X)$.
The experimental data, $F_{\rm exp}$, usually contains
the uncertainty that stems from the measurement conditions and the apparatus, so 
it is desirable for a data analysis method also to provide some information on  this associated uncertainty.
Here we focus on the inverse problem in which 
the characteristic, $F$, can be calculated from theory as a forward problem
with the function of the target variables $X$
 ($F_{\rm cal}=F_{\rm cal}(X)$).
The data analysis is then reduced 
to the optimization process 
to minimize the residual difference $R(X)$ between 
the calculated characteristics, $F_{\rm cal}(X)$, and
the experimental data, $F_{\rm exp}$, 
\begin{eqnarray}
R = R(X) \equiv | F_{\rm cal}(X) - F_{\rm exp}|. 
\label{EQ-R-FACTOR-DEF}
\end{eqnarray}
The function $R$ is called the reliability factor or R-factor in certain fields. 

Recently, we developed  data analysis software with an optimization procedure \cite{TANAKA2020_ACTA_PHYS_POLO, TANAKA_2020_Preprint} for total-reflection high-energy positron diffraction (TRHEPD) 
\cite{HUGENSCHMIDT_2016_SurfSciRep_rev, Fukaya_2018_JPHYSD, 
Fukaya_2019_Book_SurfaceStructure, Mochizuki_2016_PCCP_TiO2, ENDO_20208_Carbon}, 
a novel experimental technique for the accurate determination of surface structure. Here, the target variable set $X$ typically consist of the atomic positions on the topmost surface layer and sub-surface layers below. Before the software was developed, the determination of $X$ was performed by using a trial-and-error approach without a systematic optimization algorithm. In order to make the process more objective, we first developed an automatic optimization software for the surface atomic positions \cite{TANAKA2020_ACTA_PHYS_POLO}. Then we applied a two-stage optimization procedure \cite{TANAKA_2020_Preprint}, in which the first stage is a grid-based global search of the candidate regions in the $X$ space where the absolutely optimized $X$ could possibly exist, thus avoiding the possibility of ending up with a local optimization. The second stage is a local search for a final solution. 

The present article demonstrates the first application of our software to the analysis of real experimental data for  
Si$_4$O$_5$N$_3$ / 6H-SiC (0001)-($\sqrt{3} \times \sqrt{3}$) R30$^\circ$
\cite{Shirasawa_PRL_2007_SION,SHIRASAWA2009_PRB_SiON,Mizuno_SurfSci_2017_SiON} surface. 
The surface structure  consists of atoms to sub-nanometer depth.
We used an optimization analysis  for the determination of the surface structure 
and, then, applied a sensitivity analysis,
a method of selecting an appropriate variable set $X$ to which the residual difference $R(X)$ is sensitive. 
This may offer benefits since data analysis in a large data-space dimension, $n = {\rm dim}(X)$, may impact on reliability and/or incur high computational cost.

The present article is organized as follows:
an overview of TRHEPD is described in Sec.~\ref{SEC-TRHEPD};
the present experiment is detailed in Sec.~\ref{SEC-EXPERIMENT};
the method and result of the data analysis are shown in 
Sec.~\ref{SEC-DATA-ANALYSIS}; 
and a summary is given in Sec.~\ref{SEC-SUMMARY}. 

\section{Overview of TRHEPD \label{SEC-TRHEPD}}

TRHEPD was first proposed in 1992 by Ichimiya \cite{ichimiya1992_TRHEPD} and realized in a study in 1998 by Kawasuso and Okada \cite{kawasuso1998_prl}.
Following a period of initial development by the Kawasuso group, this technique has been actively progressed in the last decade
at the Slow Positron Facility (SPF), 
Institute of Materials Structure Science (IMSS), High Energy Accelerator Research Organization (KEK)
\cite{HUGENSCHMIDT_2016_SurfSciRep_rev, Fukaya_2018_JPHYSD, 
Fukaya_2019_Book_SurfaceStructure, Mochizuki_2016_PCCP_TiO2, ENDO_20208_Carbon}.
Since the volume fraction of the surface region is much smaller than that of the  bulk region,
the experimental technique should be selectively sensitive to the atoms in the surface region. 
While the experimental setup of TRHEPD is essentially the same as that for reflection high-energy electron diffraction (RHEED),  
TRHEPD has a higher surface sensitivity than RHEED due to a particular physical property of materials: 
since the electrostatic potential in every material is positive, 
the potential energy of the positron in the material is positive while that of the electron is negative.
Consequently, the positron diffraction technique is more suitable for the structural analysis of topmost and sub-surface atomic layers.
This makes the positron an ideal probe of surface structure.
In fact, the measuring depth of the positron is on the sub-nanometer order, 
as shown through a model calculation in Fig.~3 of Ref.~\cite{Fukaya_2018_JPHYSD}, which leads to the surface selectivity of TRHEPD and RHEED  being remarkably different  
in the lower incident glancing-angle region. 

The  target variables $X$ in TRHEPD is  
the surface structure or 
a set of the atomic positions $(X_1, X_2, ....,X_n)$ on the topmost and  sub-surface layers below. 
The expected diffraction characteristics from the atomic arrangements in the probed surface region, 
called a rocking curve, is calculated
as a function of the surface structure 
 ($F_{\rm cal} \equiv F_{\rm cal}(X)$), like that in RHEED, 
by solving the partial differential equation in a fully-dynamical quantum diffraction theory
\cite{Ichimiya_1983_JJAP_RHEED_solver, ICHIMIYA_1987_SurfSciLett_OneBeam}.   
In the previous article \cite{TANAKA_2020_Preprint},
we used the diffraction data of Ge(001)-c($4 \times 2$) surface generated numerically by the partial differential equation, instead of real experimental data.
The present article reports, first, 
the analysis of real experimental data with the additional sensitivity analysis.

\section{Experiment \label{SEC-EXPERIMENT}}

\subsection{Material}

\begin{figure}[h]
\begin{center}
  \includegraphics[width=5.5cm]{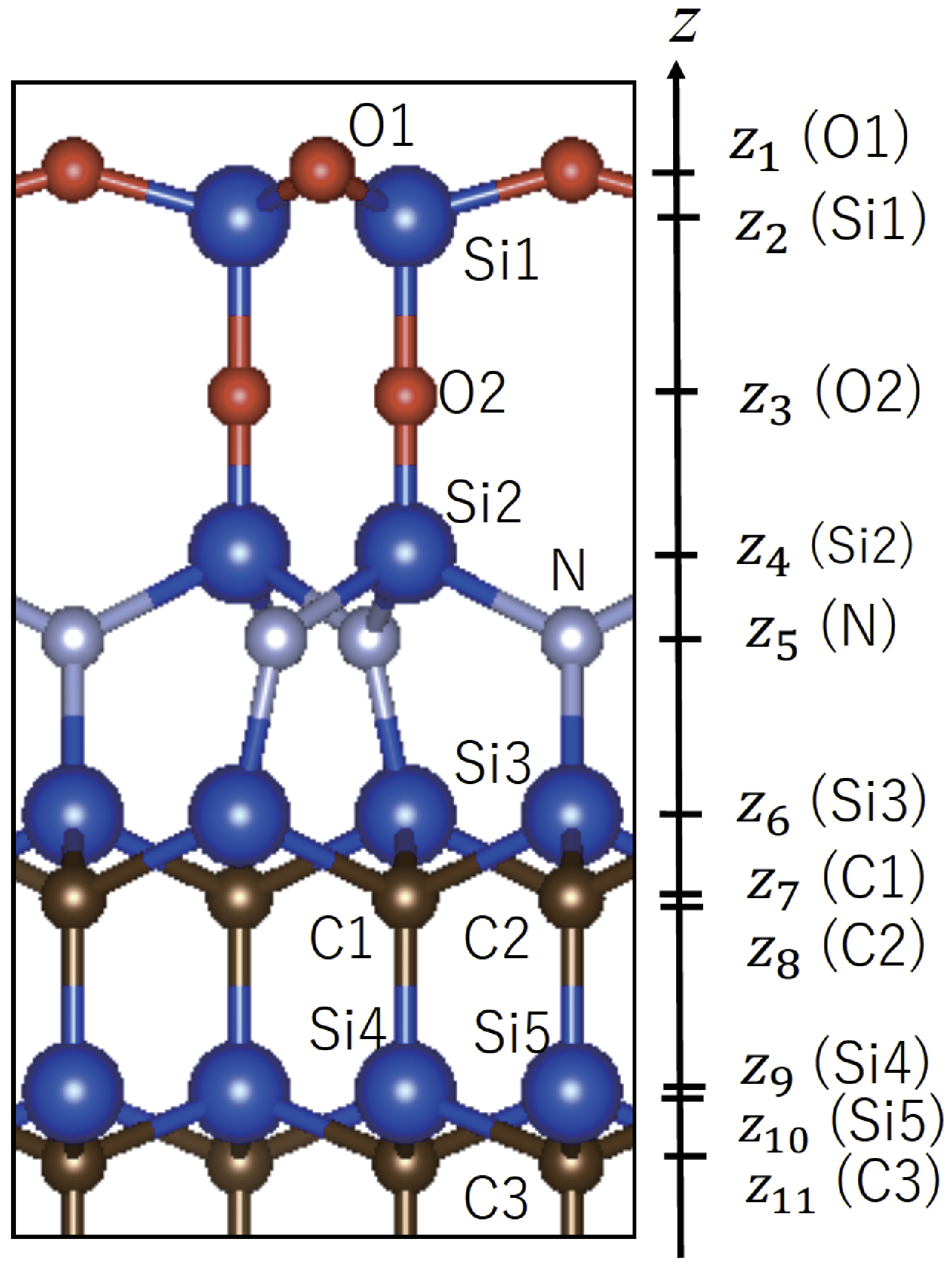}
\end{center}
\caption{Side view of the surface structure of
Si$_4$O$_5$N$_3$ / 6H-SiC (0001)-($\sqrt{3} \times \sqrt{3}$) R30$^\circ$
\cite{Shirasawa_PRL_2007_SION,SHIRASAWA2009_PRB_SiON,Mizuno_SurfSci_2017_SiON}.
}
\label{FIG-SION-STRUCTURE}       
\end{figure}

The surface measured by TRHEPD in the present work is
Si$_4$O$_5$N$_3$ / 6H-SiC (0001)-($\sqrt{3} \times \sqrt{3}$) R30$^\circ$ 
\cite{Shirasawa_PRL_2007_SION,SHIRASAWA2009_PRB_SiON,Mizuno_SurfSci_2017_SiON}.
Figure \ref{FIG-SION-STRUCTURE} shows 
a side view of the structure reported in 
references ~\cite{Shirasawa_PRL_2007_SION,SHIRASAWA2009_PRB_SiON,Mizuno_SurfSci_2017_SiON}.
The $z$ axis is chosen to be perpendicular to the surface.
The top view is found, for example,
in Fig.1(c) of
Ref.\cite{Mizuno_SurfSci_2017_SiON}.
We restricted ourselves to the analysis of the TRHEPD data in the one-beam condition (see Sec.~\ref{SEC-MEASUREMENT}) 
which is sensitive only to the atomic coordinates perpendicular to the surface, $z$. 
The notation of the atomic sites 
in Fig. \ref{FIG-SION-STRUCTURE}, such as O1, Si1 and O2,
follows that in Ref.~\cite{Mizuno_SurfSci_2017_SiON}.
The $z$ coordinates of the sites are denoted as 
$z_1({\rm O1})$, $z_2({\rm Si1)}$, $z_3({\rm O2)}$, $z_4({\rm Si2)}$, $z_5({\rm N)}$, $z_6({\rm Si3)}$, 
$z_7({\rm C1)}$, $z_8({\rm C2)}$, $z_{9}({\rm Si4)}$, $z_{10}({\rm Si5)}$, and $z_{11}({\rm C3)}$
in the descending order ($z_1 > z_2 > .... > z_{11}$).

The preparation of the sample 
(of size $5 \times 10 \, {\rm mm}^2$) 
was as follows. 
The epitaxial SiON layer was grown on an on-axis
4H-SiC(0001) substrate \cite{Shirasawa_PRL_2007_SION}. The substrate was initially etched with hydrogen gas under atmospheric pressure at 1623K for 15 min to clean and atomically smoothen the surface \cite{NAKAGAWA2003_PRL_SiC}, and subsequently exposed to nitrogen gas at the same temperature. Oxygen gas was not intentionally introduced in this experiment but may have been mixed in the reactant as an impurity. 
Samples were then transferred to the TRHEPD measurement chamber and
further annealed (973 K, 60 min) in UHV (less than $1.0 \times 10^{-7}$ Pa) to remove surface contamination.

\subsection{TRHEPD measurement \label{SEC-MEASUREMENT}}

The details of the TRHEPD station at the KEK-IMSS-SPF are described elsewhere \cite{Mochizuki_2016_PCCP_TiO2, WADA_2012_EuroPhysD, MAEKAWA_2014_EuroPhysD} and 
a brief overview is given below. 
A linac-based brightness-enhanced positron beam with an energy of 10 keV was used. 
The diffraction patterns were obtained under the one-beam condition \cite{ICHIMIYA_1987_SurfSciLett_OneBeam, Fukaya_2018_JPHYSD},
where the beam azimuthal angle was set at 7.5$^{\circ}$ off the $[11\bar{2} 0]$ direction.
In the one-beam condition, the diffraction spot intensity depends primarily on the atomic coordinates 
perpendicular to the surface because the in-plane diffraction is effectively suppressed. 
The dependence of the 00-spot diffraction intensity ($I_{00}^{\rm (exp)}(\theta)$)
on the glancing angle ($\theta$) of incidence, called the rocking curve \cite{ICHIMIYA_1987_SurfSciLett_OneBeam, Fukaya_2018_JPHYSD}, was extracted from the series of TRHEPD patterns acquired, taken with an exposure time of 1.5 min each. 
The glancing angle was varied from $\theta_{\rm min}=0.5^{\circ}$ to $\theta_{\rm max}=6.5^{\circ}$ 
in $\Delta \theta = 0.1^{\circ}$ steps by tilting the sample.
The experimental rocking curve was expressed as 
$F_{\rm exp} = (I^{\rm (exp)}_{00}(\theta_1=\theta_{\rm min}),  ..., I^{\rm (exp)}_{00}(\theta_\nu=\theta_{\rm max}))$,
with $\nu \equiv 1+(\theta_{\rm max}-\theta_{\rm min})/\Delta \theta =61$.
The experimental data and the corresponding calculated data were normalized as 
$|F_{\rm exp}|= \{ \sum_i^\nu |I^{\rm (exp)}_{00}(\theta_i)|^2\}^{1/2} = 1$
and $|F_{\rm cal}|= \{ \sum_i^\nu |I^{\rm (cal)}_{00}(\theta_i)|^2\}^{1/2} = 1$, respectively,
when the R-factor (\ref{EQ-R-FACTOR-DEF}) was calculated. 
Since only the data obtained in the one-beam condition were analyzed, the procedure described 
in the next subsection concerns only that with the $z$ coordinates $(X=(z_1, z_2, ...))$.

\section{Data analysis  \label{SEC-DATA-ANALYSIS}} 

This section describes the analysis of 
TRHEPD data of 
Si$_4$O$_5$N$_3$ / 6H-SiC (0001)-($\sqrt{3} \times \sqrt{3}$) R30$^\circ$.
The analysis consists of
the auto-optimization procedure described in 
Refs.~\cite{TANAKA2020_ACTA_PHYS_POLO, TANAKA_2020_Preprint}
and the sensitivity analysis. 
Among existing papers of TRHEPD 
(\cite{Fukaya_2018_JPHYSD} and references therein), 
the structure $X$ is accepted as a final solution 
when the R-factor ($R$) is optimized to be
less than 0.02 ($R(X) \le 0.02$).

Candidates for the atomic positions of the present sample are found in Refs.~\cite{Shirasawa_PRL_2007_SION,Mizuno_SurfSci_2017_SiON}. 
In Ref. \cite{Shirasawa_PRL_2007_SION},
the atomic positions of Si$_4$O$_5$N$_3$ / 6H-SiC (0001)-($\sqrt{3} \times \sqrt{3}$) R30$^\circ$ surface were determined from the LEED experiment by assuming p3 symmetry.
Later, in Ref. \cite{Mizuno_SurfSci_2017_SiON}, 
they were determined from the LEED experiment by assuming p31m symmetry,
a higher degree of symmetry than p3 symmetry.
The atomic positions in the two papers are similar.
In the present analysis of the iterative local optimization, 
the initial $z$ coordinates were chosen to be
equivalent to those in Table I of Ref.~\cite{Mizuno_SurfSci_2017_SiON}; 
 ($z_1$, $z_2$, $z_3$, $z_4$, $z_5$, $z_6$, $z_7$, $z_8$, $z_9$, $z_{10}$) =
($z_1^{\rm (ini)}$, $z_2^{\rm (ini)}$, $z_3^{\rm (ini)}$, $z_4^{\rm (ini)}$, $z_5^{\rm (ini)}$, $z_6^{\rm (ini)}$, 
$z_7^{\rm (ini)}$, $z_8^{\rm (ini)}$, $z_9^{\rm (ini)}$, $z_{10}^{\rm (ini)}$)
$=$ (9.19\AA, 8.67\AA, 7.04\AA, 5.45\AA,  4.83\AA, 3.11\AA,  2.63\AA,  2.44\AA, 0.67\AA, 0.59\AA),
where the coordinate
$z_{11}$ 
in Fig.~\ref{FIG-SION-STRUCTURE} 
is set to be the origin ($z_{11} \equiv 0$).
The atomic positions at the deeper layers $(z < z_{11})$ 
are set to be that in the bulk. 

The present calculations were carried out by the supercomputer Oakforest-PACS 
with Intel Xeon Phi$^{\rm TM}$ 7250 processors. 
The use of the supercomputer was crucial only in the sensitivity analysis,
since the computational cost of the optimization analysis is very small.
The calculated rocking curve, $F_{\rm cal}=F_{\rm cal}(X)$, was generated by 
the solver routine of the fully-dynamical quantum diffraction theory
used in Ref.~\cite{HANADA_PRB_1995},
as in our previous works  \cite{TANAKA2020_ACTA_PHYS_POLO, TANAKA_2020_Preprint}.

\subsection{Optimization analysis with eight variables \label{SEC-ANALYSIS-OPT} }

The optimization procedure was carried out 
with the R-factor as the function of 
the eight coordinates, $\{z_i\}_{i=1,8}$
($R=R(z_1, z_2, z_3, z_4, z_5, z_6, z_7, z_8)$.
The eight coordinates were chosen so that
the probed region consists of the whole SiON region ($z_1, z_2, z_3, z_4, z_5$) 
and a set of bulk SiC layers ($z_6, z_7, z_8$).
The variables $z_9, z_{10}, z_{11}$  are fixed to be those in the initial structure, which are similar to the bulk positions.
The calculated system contains a semi-infinite bulk region,
as in the existing papers of TRHEPD measurement 
(\cite{Fukaya_2018_JPHYSD} and references therein).

The iterative optimization procedure was made using 
the gradient-free, Nelder-Mead algorithm
\cite{NMmethod-a, NMmethod-b}
for which a Python code was developed in the previous work
\cite{TANAKA2020_ACTA_PHYS_POLO, TANAKA_2020_Preprint}.
The Nelder–Mead algorithm was performed by a module in the scipy library (scipy.optimize.fmin).
Hereafter we use the notation $X$ specifically for the vector 
$X=(z_1, z_2, z_3, z_4, z_5, z_6, z_7, z_8)^{\rm T}$
in the eight-dimensional data space. 
In the Nelder-Mead algorithm,
some of the set of $n+1(=9)$ sampling points $\{X^{(l)}\}^{[k]}(l=0,1...,8)$ are 
replaced by new points suitably found by calculation in every iteration, 
where 
$l$ is the sampling point index and 
$k$ is the iteration step index ($k=0,1,2...$).
The best sampling point $ X^{[k]} \equiv {\rm argmin}_l(\{R(X^{(l)})\}^{[k]}))$ 
is obtained in the course of  the iteration.
The iterative procedure is performed 
until the R-factor value converges 
within a given  criteria of $\Delta R = 5 \times 10^{-4}$.
The criteria value $\Delta R = 5 \times 10^{-4}$ is lower than the required threshold of magnitude,
since an R-factor value less than 0.02  ($R(X) \le 0.02$) is regarded as acceptable  in the TRHEPD experiments noted at the beginning of the present section. 
Such a criteria value was used so as to demonstrate that
even such a narrow convergence can be attained 
within short computational time. 
The initial data of the sampling points $\{X^{(l)} \}^{[k]} (l=0,1,2,...,8)$ is chosen by the researcher.
The structure in Ref.~\cite{Mizuno_SurfSci_2017_SiON} was chosen 
to be the initial data of the zero-th sampling point 
$X^{[0](0)} =(z^{[0](0)}_1, z^{[0](0)}_2, ..., z^{[0](0)} _8)^{\rm T}$, as mentioned above. 
The initial data of the other sampling points $\{ X^{(l)}\}^{[0]} (l=1,2,...,8)$ were chosen so that 
the $l$-th sampling point 
$X^{(l)[0]} =(z^{(l)[0]}_1, z^{(l)[0]}_2, ..., z^{(l)[0]} _8)^{\rm T}$ 
is displaced by $0.05$ \AA $\,$ from 
the zero-th sampling point $X^{[0](0)}$ only in the $l$-th coordinate   
($z^{(l)[0]}_l= z^{(0)[0]}_l + 0.05$\AA,
$z^{(l)[0]}_i = z^{[0](0)}_i$ for $i \ne l$).

The optimization procedure with the  Nelder-Mead algorithm 
converged at the 42-th iteration. 
This took the computational time, $T$, of approximately
one minute of one CPU of the Oakforest-PACS. 
The R-factor values for the initial and converged structures 
were $R^{\rm (ini)} = R(X^{\rm (ini)}) = 2.28 \times 10^{-2}$ and
$R^{\ast} = R(X^{\ast}) = 0.91 \times 10^{-2}$, respectively.
The converged point was
$X^{\ast} = (z_1^{\ast}, z_2^{\ast}, ..., z_8^{\ast})
\approx$ (9.09\AA, 8.67\AA, 7.03\AA, 5.60\AA, 4.80\AA, 3.11\AA, 2.64\AA, 2.44\AA).
The difference between the initial and converged structures
($\delta z_i \equiv z_i^{\ast} - z_i^{\rm (ini)}$) was 
($\delta z_1$, $\delta z_2$, $\delta z_3$, $\delta z_4$, $\delta z_5$, $\delta z_6$, $\delta z_7$, $\delta z_8$)
$\approx$ 
(-0.10\AA, 0.00\AA, -0.01\AA, 0.15\AA, -0.03\AA, 0.00\AA, 0.01\AA, 0.00\AA).

\begin{figure}[h]
\begin{center}
  \includegraphics[width=7cm]{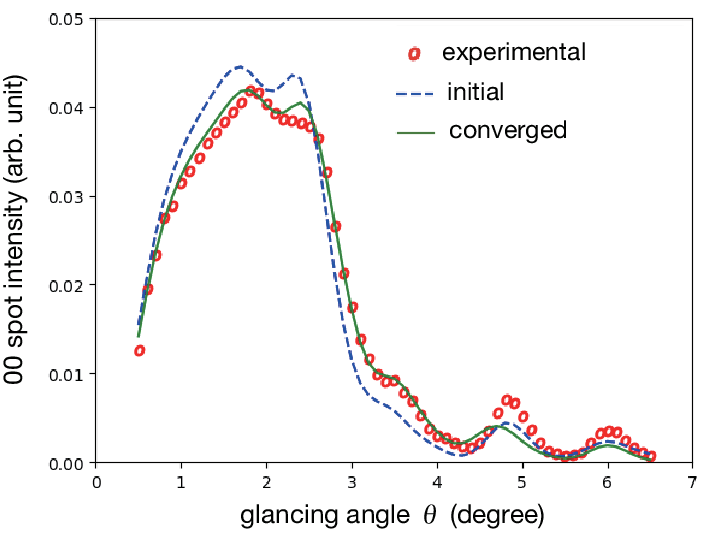}
\end{center}
\caption{
Comparison of the experimental rocking curve (open circles), the calculated values for the initial structure (dashed line) and the calculated values for the converged structure (solid line) in the eight-variable optimization.
}
\label{FIG-OPTIMIZATION}       
\end{figure}

\begin{figure}[ht]
\begin{center}
  \includegraphics[width=5cm]{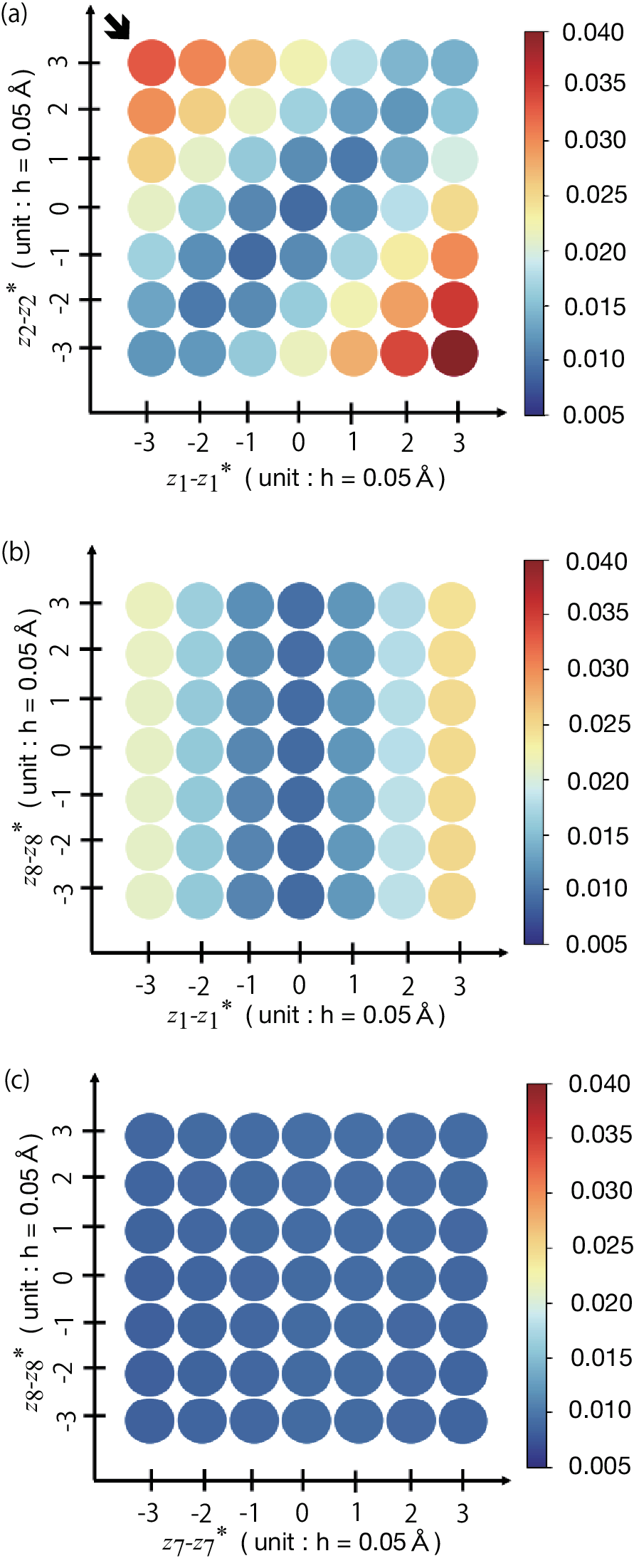}
\end{center}
\caption{(a)  The color-coded isovalue plots of $R(X)$ on the local grid of the $z_1$-$z_2$ plane.
The $(1, -1, 0, 0, 0, 0, 0, 0)^{\rm T}$ direction is depicted as the bold arrow located at the left upper corner. 
(b)  The color-coded isovalue plots on the local grid of the $z_1$-$z_8$ plane.
(c)  The color-coded isovalue plots on the local grid of the $z_7$-$z_8$ plane.
The difference between the maximum and minimum  values of $R(X)$
among the plotted grid points is less than $0.1$ in (c).  
}
\label{FIG-COLOR-MAP}       
\end{figure}

Figure~\ref{FIG-OPTIMIZATION} shows 
the calculated rocking curves $F_{\rm cal}$ in the initial and converged structures, 
together with the experimental data $F_{\rm exp}$. 
The calculated rocking curves $F_{\rm cal}$ in the converged structure agrees with the experimental data $F_{\rm exp}$
more closely than the rocking curve of the initial structure. 
Note that TRHEPD technique is sensitive to the small difference ($|\delta z_i| \le 0.15$\AA). 
While it is certain that the structure of the present specimen is
essentially the same as that in the previous papers
\cite{Shirasawa_PRL_2007_SION,SHIRASAWA2009_PRB_SiON,Mizuno_SurfSci_2017_SiON},
it would be of interest to see whether the variation in the sample preparation procedure produced this small difference.

\subsection{Sensitivity analysis \label{SEC-ANALYSIS-UNCERTAINTY}}

The sensitivity analysis was carried out after the optimization, 
by capturing the numerical behavior of $R(X)$ near the converged point $X^{\ast}$.
We used supercomputers to compute $R(X)$ 
on the local grid of the coordinates $z_i \equiv z_i^{\ast} + m h$ for $m=-3, -2, -1, 0, 1, 2, 3$ and $i=1, ... ,8$
with the uniform grid interval $h=0.05$ \AA.
The total number of eight-dimensional grid points was $N_{\rm grid} = 7^8 =5,764,801$.
Although the total operational cost on the grid points is large, 
a fast computation is possible on modern massive parallel supercomputers.  
The parallelism was carried out with Message Passing Interface (MPI) and
each CPU executed one MPI process.
The total computational time $T$ with the $N_{\rm grid}$ grid points was 
approximately 1.5 hours
using $N_{\rm CPU}=2,048$ CPUs of Oakforest-PACS.
Each CPU calculated the R-factor $R(X)$ at approximately 
$N_{\rm grid}/ N_{\rm CPU} \approx 2815$ grid points.

Figure ~\ref{FIG-COLOR-MAP} demonstrates the anisotropic sensitivity
among a few two-dimensional isovalue 
plots of $R=R(z_1, z_2, ..., z_8)$ on the local grid with the use of color-coding. 
Figure ~\ref{FIG-COLOR-MAP}(a) is the isovalue plot
on the $z_1$-$z_2$ plane,
where all the other variables are fixed to be the converged values
($z_i = z_i^{\ast}$ for $i=3-8$).
We found that the function $R(X)$ is quite sensitive to the deviation in
the $(1,-1,0,0,0,0,0,0)^{\rm T}$ direction, being proportional to the distance of the two atomic
layers $(z_1-z_2)$, therefore showing that there is a significant contribution to
the diffraction signal by the interaction of the positron waves
scattered from the two atomic layers $z_1$ and $z_2$.
Figure ~\ref{FIG-COLOR-MAP}(b)
is the isovalue plot on the $z_1$-$z_8$ plane
showing that the function $R(X)$ is insensitive to 
the deviation in the $(0, 0, 0, 0, 0, 0, 0, 1)^{\rm T}$ direction or the $z_8$-axis direction. 
Figure ~\ref{FIG-COLOR-MAP}(c)
is the isovalue plot on the $z_7$-$z_8$ plane
and indicates that the function $R(X)$ is 
insensitive to any deviation on the $z_7$-$z_8$ plane.
These properties are consistent with  surface selectivity or the statement that
the TRHEPD measurement observes mainly the shallow region with $z \ge z_6$.

The anisotropic behavior of the function $R(X)$ 
was systematically examined 
by the variance-covariance matrix $S$,
whose $(i,j)$ component is defined as 
\begin{eqnarray}
S_{ij} \equiv \frac{1}{\Omega} \int (z_i - z_i^{\ast}) (z_j - z_j^{\ast}) W(X) d X
\label{EQ-S}
\end{eqnarray}
for $i,j=1,2,...,8$ with
the weight function 
\begin{eqnarray}
W(X) \equiv {\rm e}^{- (R(X)-R^{\ast})/s} \label{EQ-KERNEL}
\end{eqnarray}
and the normalization factor
\begin{eqnarray}
\Omega \equiv  \int  W(X) d X.  \label{EQ-C}
\end{eqnarray}
A parameter $s=0.002$, called the scaling parameter, is introduced,
which is the tolerance measure of the uncertainty for the R-factor value.

It is noteworthy that 
the function $R(X)$ can be written formally
by the second-order Taylor expansion  
\begin{eqnarray}
R(X)/s &\approx& R(X^{\ast})/s \nonumber \\
 &+& \sum_{i,j}^n(z_i - z_i^{\ast})  \Gamma_{ij}  (z_j - z_j^{\ast}),
 \label{EQ-TAYLOR}
\end{eqnarray}
with $\Gamma_{ij} \equiv \partial^2 (Rs^{-1}) / (\partial z_i \partial z_j)$ of  the curvature matrix $\Gamma$.
If the third- and higher-order terms of the Taylor expansion are ignored,
the inverse matrix of $S$ is the curvature matrix $\Gamma$ ($S = \Gamma^{-1}$).
The isovalue plot of Eq.~(\ref{EQ-TAYLOR}) ($R(X)={\rm (constant)}$) forms an $n$-dimensional ellipsoid and
the $k$-th eigenvector $\bm{v}_k$ indicates a principal axis of the $n$-dimensional ellipsoid.

When the integrals in Eq.~(\ref{EQ-S}) and  Eq.~(\ref{EQ-C}) 
are reduced to the sum over the local grid points 
defined at the beginning of this section,
the matrix $S$ was obtained as
\begin{eqnarray}
& &  \hspace{-1cm} S \approx \nonumber \\
& & \hspace{-1cm} \left(
\begin{array}{cccccccc}
7.39 & 6.59 & 3.38 & 3.20 & 0.43 & -0.02 &  0.04 & 0.19 \\
         & 7.52 & 3.18 & 2.58 & 1.30 &  0.33 & -0.47 & 0.12 \\
         &          & 8.18 & 1.07 & 0.23 &  0.53 &  0.02 & 0.08 \\
         &          &          & 7.59 & 2.04 &  1.41 & -0.40 & 0.13 \\
         &          &          &          & 9.02 &  0.63 & -0.05 & 0.17 \\
         &          &          &          &          &  8.93 & -1.06 & 0.05 \\
         &          &          &          &          &           &  10.66 & 0.01 \\
         &          &          &          &          &           &             & 9.87 \\
\end{array}
\right)  \nonumber \\
\label{EQ-S-VALUE}
\end{eqnarray}
 in units of $10^{-3}$ \AA$^2$.  
The matrix, $S$, is symmetric and 
Eq.~(\ref{EQ-S-VALUE}) shows 
only the upper triangular elements explicitly. 
The importance of the off-diagonal elements is characterized by the quantity
$q_ i \equiv (\sum_{j\ne i}|S_{ij}|) / |S_{ii}|$ 
for each column ($i=1,2,...,8$). 
The values were $q_1 = 1.87, q_2=1.94, q_3=1.04, q_4=1.43, q_5=0.54, q_6=0.45, 
q_7=0.19, q_8 = 0.08$. 
It was found that
the off-diagonal elements are significant 
among the first to fifth columns 
($q_i \ge 0.5$ for $i=1, ...,5$).
The presence of the significant off-diagonal elements in S indicates that the effect of the displacement of each atomic layer correlates significantly to the value of the R-factor in the first to fifth layers.

The principal deviation direction in the eight-dimensional data space 
is obtained by solving the $8 \times 8$ matrix eigenvalue equation
\begin{eqnarray}
S \bm{v}_k = \lambda_k \bm{v}_k
\label{EQ-EIG}
\end{eqnarray}
with $0 \le \lambda_1 \le \lambda_2 \le ... \le \lambda_8$ and $|\bm{v}_k|=1$.
The $k$-th eigenvalue $\lambda_k$ indicates the directional variance in the $\bm{v}_k$ direction. 
The values were
$\lambda_1 = 0.73 \times 10^{-3} $\AA$^2$, 
$\lambda_2 = 4.53 \times 10^{-3} $\AA$^2$, 
$\lambda_3 = 6.18 \times 10^{-3} $\AA$^2$, 
$\lambda_4 = 8.20 \times 10^{-3} $\AA$^2$, 
$\lambda_5 = 9.62 \times 10^{-3} $\AA$^2$, 
$\lambda_6 = 9.92 \times 10^{-3} $\AA$^2$, 
$\lambda_7 = 11.3 \times 10^{-3} $\AA$^2$, 
$\lambda_8 = 18.7 \times 10^{-3} $\AA$^2$.
The sum of the eigenvalues
${\rm Tr}[S] = \sum_{k=1}^8 \lambda_k$ quantifies the total uncertainty and
the difference between the eigenvalues indicates the anisotropy in the uncertainty.
For example, 
the first eigenvector $\bm{v}_1$ with the smallest eigenvalue (directional variance) ${\lambda_1}$
represents  the most sensitive direction or the direction with the smallest uncertainty.

\begin{figure}[h]
\begin{center}
  \includegraphics[width=7.5cm]{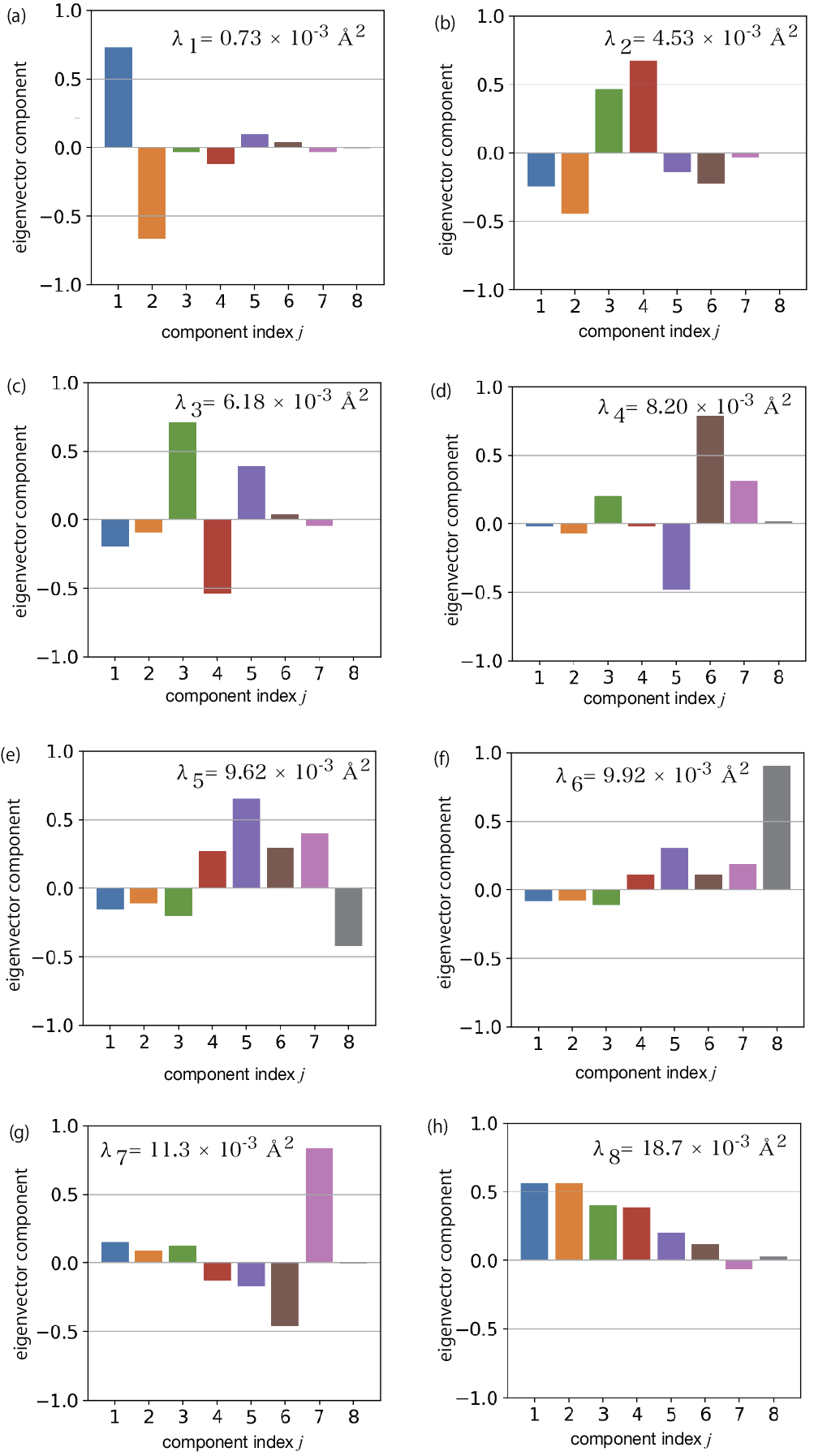}
\end{center}
\caption{
The components of 
the eigenvectors
$\bm{v}_k = (v_{1k}, v_{2k}, .,v_{ik},., v_{8k})^{\rm T}$  for
(a)$k=1$, (b)$k=2$, (c)$k=3$, (d)$k=4$, (e)$k=5$, (f)$k=6$, (g)$k=7$, (h)$k=8$.
The values of  the eigenvalues (directional variance) $\lambda_k$ ($k=1,2,...,8$) are given in each graph. 
}
\label{FIG-EIG-VECTOR}       
\end{figure}

Figure \ref{FIG-EIG-VECTOR} shows the values of 
the eigenvector components $\bm{v}_k = (v_{1k}, v_{2k}, .., v_{8k})^{\rm T}$ for $k=1,2,...,8$
revealing the principal axis directions for the anisotropic sensitivity. 
As an example, 
the first eigenvector $\bm{v}_1$ is approximated to be 
$\bm{v}_1 \approx (1/\sqrt{2})(1,-1,0,0,0,0,0,0)^{\rm T}$ and thus 
the function $R(X)$ should be quite sensitive to the deviation in the $(1,-1,0,0,0,0,0,0)^{\rm T}$,
as seen on Figure ~\ref{FIG-COLOR-MAP}(a).
As another example, the sixth and seventh eigenvectors are similar to 
the eighth and seventh original axis vectors
($\bm{v}_6 \approx (0, 0, 0, 0, 0, 0, 0, 1)^{\rm T}, 
\bm{v}_7 \approx (0, 0, 0, 0, 0, 0, 1, 0)^{\rm T}$), respectively, 
which indicates that 
almost no interacting wavefunction is formed
by the scattering process at the seventh and eighth atomic layers and so
the variables $z_7$ and $z_8$ are barely correlated to the other variables.
Thus the function $R(X)$ may be almost unchanged 
on the $(z_7, z_8)$ plane, as seen on Figure ~\ref{FIG-COLOR-MAP}(c). 
In addition, 
it is also noted that $\bm{v}_8 \approx (1/\sqrt{5})(1,1,1,1,1,0,0,0)^{\rm T}$ in Figure \ref{FIG-EIG-VECTOR}(h), 
which means that 
the function $R(X)$ is hardly affected by 
the constant shift among $z_1, z_2, z_3, z_4$ and $z_5$
($z_i \Rightarrow z_i + \alpha$ for $i=1,...,5$).
In other words,
the function $R(X)$ is contributed to mainly by 
the relative positions between the atoms 
within the slab region that contains the first to fifth layers.

\subsection{Optimization analysis with six variables \label{SEC-ANALYSIS-OPT-SIX} }

The above sensitivity analysis implies that
the optimization analysis with the six variables
($z_1$, $z_2$, $z_3$, $z_4$, $z_5$, $z_6$) would give a reasonable result.
Thus, we performed the optimization analysis with these six variables.
The calculated system is a slab that consists only of the eleven layers at
$z=z_1, ...., z_{11}$ in Figure \ref{FIG-SION-STRUCTURE}, and neglecting the bulk part,
since it has been shown that the R-factor for TRHEPD seems not to be sensitive (or hardly at all)
to the atomic positions of the deeper layers ($z < z_{11}$).
The variables $z_7, z_8, z_9, z_{10}, z_{11}$ are fixed to be those in the initial structure.

\begin{figure}[h]
\begin{center}
  \includegraphics[width=7.5cm]{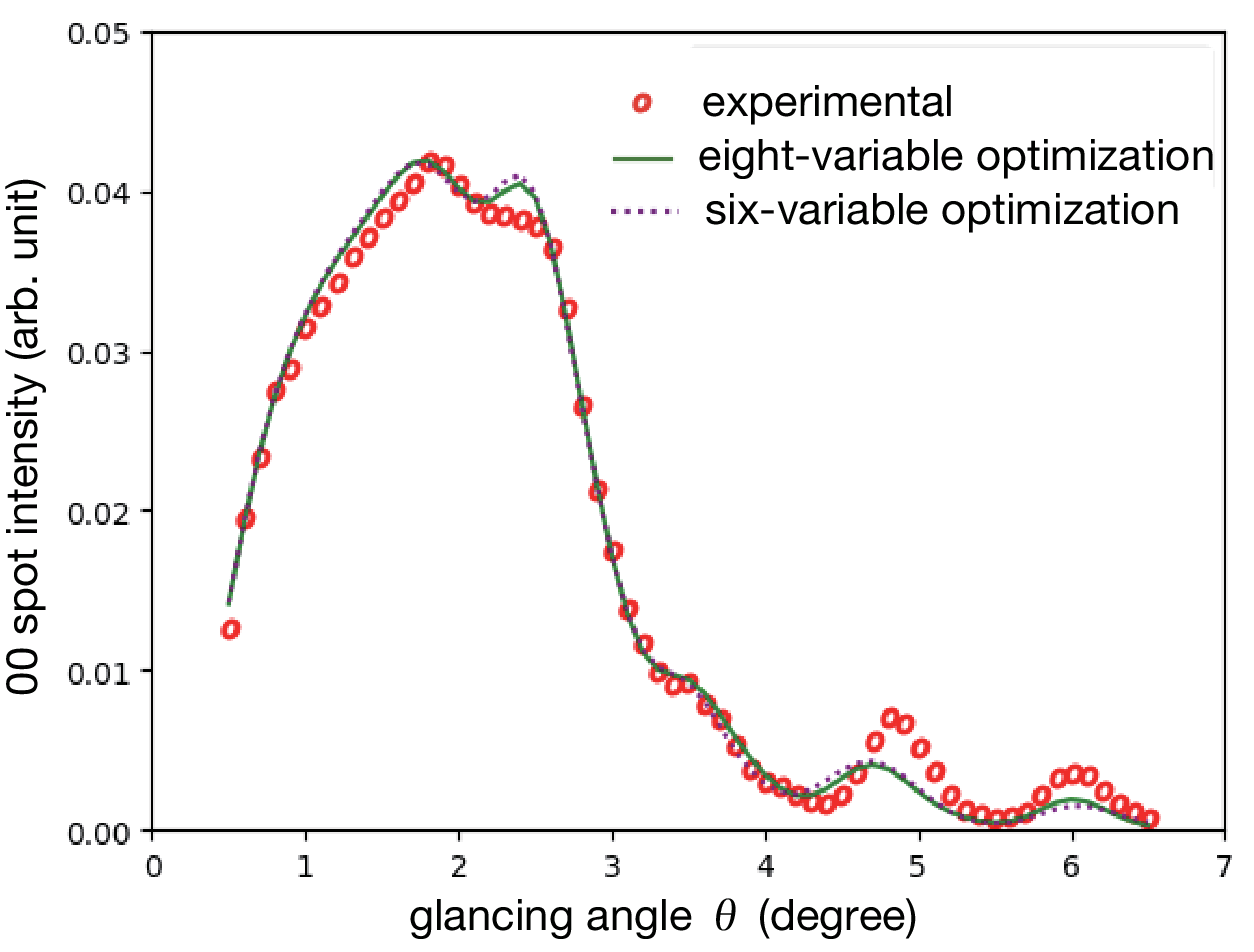}
\end{center}
\caption{
Comparison of the experimental rocking curve (open circles), the calculated values for the converged structure in the eight-variable optimization (solid line) and the calculated values for the converged structure in the six-variable optimization (dotted line).
}
\label{FIG-OPTIMIZATION_6VARIABLES}       
\end{figure}

The  optimization procedure converged at the 32-nd iteration.
The converged point was
$X^{\ast} = (z_1^{\ast}, z_2^{\ast}, ..., z_6^{\ast})
\approx$ (9.12 \AA, 8.69 \AA, 7.08\AA, 5.58\AA, 4.76\AA, 3.09\AA).
The R-factor value at the converged structure was
$R^{\ast} = R(X^{\ast}) = 0.97 \times 10^{-2}$.
The difference in the initial and converged structure
($\delta z_i \equiv z_i^{\ast} - z_i^{\rm (ini)}$) was 
($\delta z_1$, $\delta z_2$, $\delta z_3$, $\delta z_4$, $\delta z_5$, $\delta z_6$)
$\approx$ 
(-0.07\AA, 0.02\AA, 0.04\AA, 0.13\AA, -0.07\AA, -0.02\AA).
Figure~\ref{FIG-OPTIMIZATION_6VARIABLES} shows the calculated rocking curves for the converged structures by the six-variable optimization procedure (dotted line), together with that by the former eight-variable optimization procedure (solid line) and the experimental data (open circles).
The difference in the converged coordinates between the eight- and six-variable 
analyses was (-0.03 \AA, -0.02 \AA, -0.05 \AA, 0.02 \AA, 0.04 \AA, 0.02 \AA)
for  $(z_1^{\ast}, z_2^{\ast}, ..., z_6^{\ast})$.
Since 
the eight and six-variable optimization procedures
give only a small difference between the resulting
converged structure and the rocking curves, 
we concluded that both optimization procedures are acceptable.

The present sensitivity analysis gives a guide
for the valid practical choice of a set of variables, in terms of 
both computational cost and reliability.

\subsection{Generality of sensitivity analysis \label{SEC-SELECTIVITY} }

Finally, 
we comment on the generality of the above data-driven sensitivity analysis. 
The sensitivity analysis gives a foundation for the appropriate choice of the variable set $X$
by solving the eigenvalue equation of the variance-covariance matrix. 
For example,
if a material contains light and heavy atoms, 
the TRHEPD diffraction signal
tends to be more sensitive to a heavy atom than to a light atom. 
For such cases,
it is not trivial to choose an appropriate set of variables
because a heavy atom at a deeper layer may make a larger contribution to the diffraction signal
than a light atom at a shallower layer.
Also, 
the analysis method can handle variables of different physical dimension at the same time, 
such as the position and the coverage (occupation fraction) of each surface atom.
We hope to apply this method to other experiments for two-dimensional structures,
such as those using surface X-ray diffraction (SXRD),  low energy electron diffraction (LEED), 
low energy positron diffraction (LEPD)
\cite{TONG2000_LEPD,Wada2018_LEPD}.

It should be noted that 
a grid-based calculation is a rigorous global-search method but 
may incur a high  computational cost with a large data dimension.
The total computational cost of the grid-based calculation is proportional 
to the number of the grid points $N_{\rm grid}$.
We should recall that 
the calculation on the eight-dimensional grid in Sec.~\ref{SEC-ANALYSIS-UNCERTAINTY}
requires $N_{\rm grid}=7^8=5,764,801$ grid points 
and takes 1.5 hours using a quarter of the Oakforest-PACS supercomputer.
The grid-based calculation on the ten-dimensional grid
requires $N_{\rm grid}=7^{10}=282,475,249$ grid points
and is estimated to take approximately one day using the whole system.
Occupying the whole system for more than one day, however, is not usually allowed.
In such a high-dimensional case, the Monte Carlo sampling method is  promising 
both in the optimization procedure and 
the numerical integration such as Eq.~(\ref{EQ-S}). 
There are already research examples \cite{Anada_2017_JApplCrys_XRD_MC, Anada_2018_PRB_SXRD} 
in which the Monte Carlo method has been used for 
the analysis of SXRD data, though
the sensitivity analysis with the eigenvalue problem in Eq.~(\ref{EQ-EIG}) was not performed. 
Furthermore, it would be desirable to utilize the parallelizable Monte Carlo method \cite{Hukushima_2003_AIPConfProc_PAMC} 
that can efficiently use massively parallel supercomputers.

\section{Summary \label{SEC-SUMMARY}} 

The present article proposes
the data-analysis method with auto-optimization analysis and a sensitivity analysis.
The sensitivity analysis is 
based on the eigenvalue problem with the variance-covariance matrix, 
forming the foundation 
for an appropriate choice of the variables in the applied data analysis with practical reliability and moderate computational time.
The analysis was performed on the output from a total-reflection high-energy positron diffraction (TRHEPD) experiment. 
This study confirmed the high surface sensitivity of TRHEPD 
for the topmost and sub-surface atomic layers to sub-nanometer depths. 
The method is general and may be applied in wide range of experimental measurement techniques.

\section*{Acknowledgement}

The present research is supported partly by  
the Grant-in-Aid for Scientific Research (KAKENHI) from Japan Society for the Promotion of Science (19H04125, 19K12634 and 20H00581) and by
the Ministry of Education, Culture, Sports, Science and Technology
(MEXT) of Japan as a subgroup of 'Social and scientific priority issue (Creation
of new functional devices and high-performance materials
to support next-generation industries; CDMSI) to be tackled by
using post-K computer'.
Numerical computations were carried out
by the supercomputer Oakforest-PACS for 
Interdisciplinary Computational Science Program in the Center for Computational Sciences, University of Tsukuba 
and the Joint Usage/Research Center for Interdisciplinary Large-scale Information Infrastructures (Project ID: jh200045-NAH).
The numerical computation was carried out 
also at the Supercomputer Center, Institute for Solid State Physics, University of Tokyo 
and at the Academic Center for Computing and Media Studies, Kyoto University.
We thank Kazuyuki Tanaka and Takashi Hanada for fruitful discussions on the code.


\end{document}